 \documentclass[final,3p,times,onecolumn]{elsarticle}

\usepackage{amssymb}

 \biboptions{sort&compress}

\journal{Physica E}

	\usepackage{fancyheadings}
\renewcommand{\thispagestyle}[1]{} 

\begin{document}

\begin{frontmatter}



\title{Indirect coupling between localized magnetic moments in triangular graphene nanoflakes}


\author{Karol Sza{\l{}}owski}

\address{Department of Solid State Physics, Faculty of Physics and Applied Informatics, University of \L{}\'od\'z, ulica Pomorska 149/153, PL90-236 \L{}\'od\'z, Poland}
\ead{kszalowski@uni.lodz.pl}

\begin{abstract}
The indirect, charge-carrier mediated coupling between localized magnetic moments is studied for graphene nanoflakes of triangular shape and zig-zag edge. The characteristic feature of such nanoflakes is the presence of a shell of zero-energy states in the electronic spectrum. The tight-binding Hamiltonian supplemented with a Hubbard term is used for electronic structure calculations. The indirect RKKY (Ruderman-Kittel-Kasuya-Yosida) coupling energy is derived from the total electronic energy of the system in a non-perturbative way. The attention is focused on the on-site and plaquette impurities situated along the edge. The charge doping is also taken into account. It is found that the zero-energy states may give rise to a coupling mechanism which is describable by a first-order perturbation calculus and can yield robust indirect coupling of both ferro- and antiferromagnetic character, which dominates ovet the usual RKKY mechanism. The numerical results obtained emphasize the importance of the Hubbard term and the effect of the charge doping on the coupling.
\end{abstract}


\end{frontmatter}



%
\section{Introduction}
\label{intro}
Graphene, a first two-dimensional material \cite{Geim2004,Novoselov2005}, opens the route to new physics and applications. Among many others, prospects for graphene-based spintronics \cite{Wees2011} encourage extensive studies of magnetic properties of graphene and its derivatives \cite{Yazyev2010}. As a consequence, graphene nanostructures focus considerable efforts of researchers \cite{Snook}. Graphene nanoflakes behave in a quantum-dot manner, exhibiting discretization of energetic spectrum \cite{Wiessner2012}, what modifies their behaviour with respect to those expected for infinite graphene monolayers and allows for emergence of new phenomena. In particular, graphene nanoflakes of triangular shape attracted particular attention \cite{Ezawa2011,Ezawa2012,Ezawa2007,EzawaBook,Potasz2012,Potasz2011d,Potasz2010b,Zhou2012,Sheng2012,Li2012,Heiskanen2008,FernandezRossier2007,Zarenia2011,Zhang2012,Philpott2008,Uchoa2011,Silva2010}, mainly due to their peculiar electronic spectrum, containing a shell of degenerate zero-energy states leading to nontrivial magnetic properties for zigzag-edged structures. 

One of the issues related to magnetism in graphene systems is the problem of an indirect coupling between localized magnetic moments in graphene, mediated by the charge carriers \cite{Annika2010a,Annika2010b,Dugaev2006,Jiang2012,Peng2012,Power2012,Sherafati2011,Ferreira2011,Lee2012,Lee2012b,Power2012b,Power2013,Sherafati2012,Kogan2011,Kogan2012,Brey2007,Killi2011,Saaremi2007,Bunder2009,Cheianov2007,Wunsch2006,Klinovaja2012,Asgari2012,Roslyak2012}. This kind of coupling, known as Ruderman-Kittel-Kasuya-Yosida (RKKY) interaction \cite{Rudermann1954,Kasuya1956,Yosida1957}, presents an unique behaviour in graphene, due to its linear dispersion relation, unlike this found in metallic two-dimensional systems \cite{Beal-Monod1987}. Moreover, the oscillatory properties of RKKY coupling between on-site impurities in graphene are ruled by the principle relating the sign of coupling integral to the sublattice in which both magnetic impurities are placed. On the other hand, in geometrically confined graphene systems, the indirect coupling differs from that in infinite systems \cite{Szalowski2011,Szalowski2012,Bunder2012,Klinovaja2012}. The already studied systems include nanoflakes with zig-zag and armchair edges and equal number of carbon atoms in two sublattices \cite{Szalowski2011} as well as graphene nanoribbons of both kinds of edges \cite{Szalowski2012,Annika2010a,Klinovaja2012}. Let us mention that in such circumstances, some additional mechanisms of indirect exchange, dissimilar to the typical RKKY mechanism, may become operative \cite{Szalowski2011,Szalowski2012}. This fact, together with the unique electronic properties of zigzag triangular graphene nanoflakes, makes them particularly interesting systems for further studies of indirect coupling. 

Moreover, it can be mentioned that the magnetically doped nanosized objects provide the opportunity of control over the magnetic properties by means of doping with a small number of the charge carriers. This involves both diluted magnetic semiconductor quantum dots \cite{FernandezRossier2004,Qu2006} and graphene-based nanostructures \cite{Szalowski2011} and provides an additional reason for studies of such nanoscopic systems. 

The aim of the present paper is to discuss the form of an indirect coupling between the magnetic impurities in graphene nanoflakes of triangular shape and zigzag edge. The calculations are based on the tight-binding approximation with Hubbard term for the electronic structure. The charge doping of the nanostructures is taken into account.

\section{Theory}
\label{theory}

\begin{figure}
\includegraphics[scale=0.5]{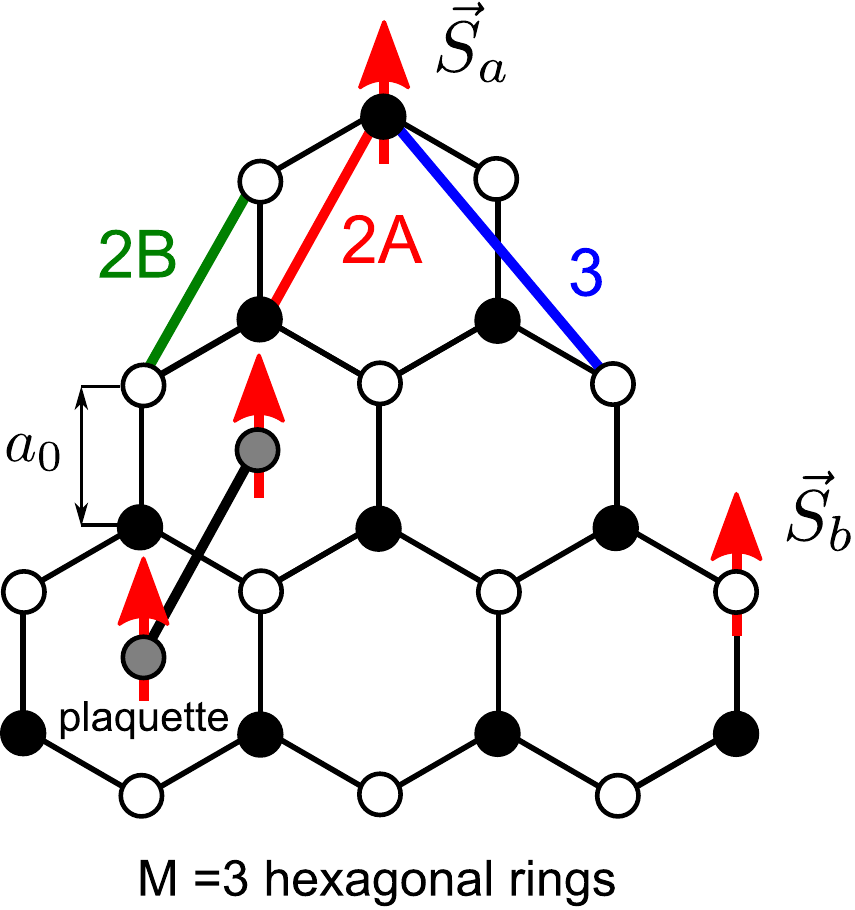} 
\caption{Schematic view of a triangular zigzag-edged nanoflake with $M=3$ hexagonal rings at each side; various impurity pairs considered in the calculations are marked; empty (filled) circles denote atoms belonging to the majority (minority) sublattice.}
\label{fig:fig1}
\end{figure}

\begin{figure}
\includegraphics[scale=1.2]{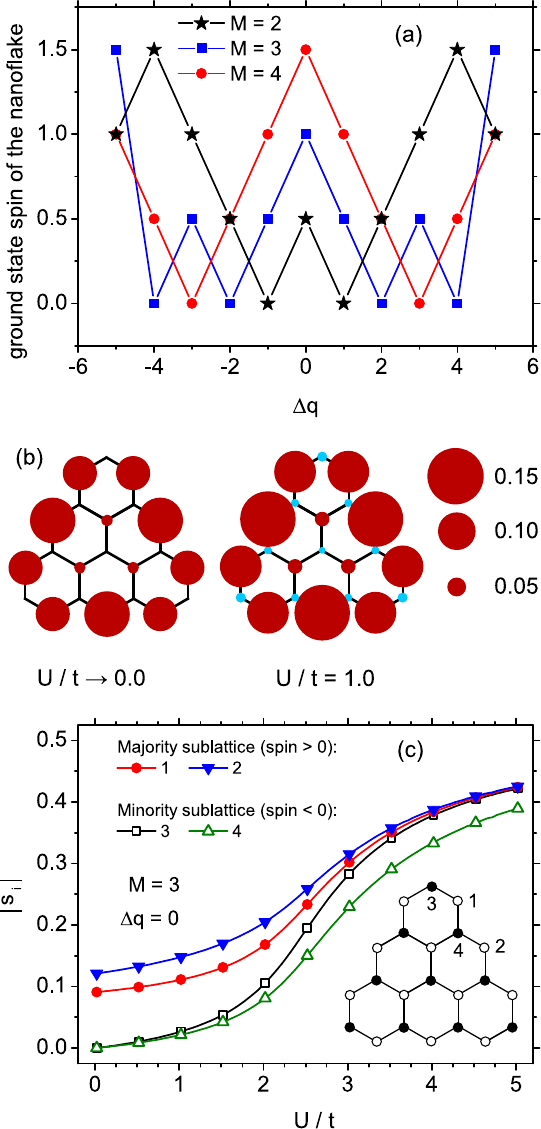}
\caption{(a) Total ground-state spin of the nanoflakes with $M=2,3,4$ as a function of charge-carrier doping; (b) total spin associated with the lattice sites of the nanoflake with $M=3$ for charge neutrality in the ground-state, for $U/t\to 0$ and $U/t=1$; (c) Absolute value of the spin density associated with various lattice sites of the $M=3$ nanoflake as a function of the normalized Hubbard on-site exchange energy $U/t$.}
\label{fig:fig1b}
\end{figure}

The subject of the study are the triangular graphene nanoflakes posessing zig-zag edge. The size of such a flake can be described using the number $M$ of hexagonal rings forming each side of the structure (see Fig.~\ref{fig:fig1}). Such GNF consists of $N=M^2+4M+1$ carbon atoms \cite{Ezawa2012}, and the two sublattices contain unequal number of atoms. Therefore, one sublattice (empty circles) is the majority sublattice, while the other sublattice (filled circles) is the minority one. The difference on the number of atoms between them amounts to $M-1$. 

In the present study, the electronic structure of the GNFs is described using the following Hamiltonian for $p^{z}$ electrons of carbon atoms \cite{Szalowski2011}: 
\begin{eqnarray}
\label{eq:eq1}
\mathcal{H}_{0}&=&-t\sum_{\left\langle i,j\right\rangle ,\sigma}^{}{\left(c^{\dagger}_{i,\sigma}c^{}_{j,\sigma}+c^{\dagger}_{j,\sigma}c^{}_{i,\sigma}\right)}-U\sum_{i}^{}{\left\langle n_{i,\uparrow} \right\rangle\left\langle n_{i,\downarrow} \right\rangle}\nonumber\\&&+\,\,U\sum_{i}^{}{\left(n_{i,\uparrow}\left\langle n_{i,\downarrow} \right\rangle+n_{i,\downarrow}\left\langle n_{i,\uparrow} \right\rangle\right)}.
\end{eqnarray}
The summation over $\left\langle i,j\right\rangle$ denotes the nearest-neighbour pairs of sites, while $\sigma=\downarrow,\uparrow$ is the electron spin. 
The total Hamiltonian involves a tight-binding nearest-neighbours hopping term with hopping integral $t$ (approximately equal to 2.8 eV \cite{CastroNeto2009}). In addition, the coulombic interactions between electrons are described by means of an on-site Hubbard term, to which a mean field approximation (MFA) is applied in a self-consistent way \cite{Wakabayashi1996,Annika2010b,Szalowski2011,Feldner2010,Yazyev2010}. Let us mention that, according to \cite{Feldner2010}, the MFA results for total energy compare successfully to exact studies of the Hubbard Hamiltonian for graphene nanostructures provided that approximately $U/t<2$, what justifies its application here.

It is worth mentioning that the value of $U$ used in such calculations does not represent an actual on-site Coulombic energy. Instead, it is an effective parameter influenced by the long-range nature of Coulombic interactions and its value for small graphene nanosystems should be close to $U/t\simeq 1$ \cite{Schuler2013}.

It follows from the studies of the electronic structure (see for example \cite{Ezawa2012,EzawaBook} or \cite{Potasz2010b,Potasz2012}) that a zigzag nanoflake with $M$ hexagons at the edge possesses a shell of $M-1$ (degenerate) zero-energy states in its spectrum. The most typical situation is that of a charge neutral, undoped nanoflake, with $\Delta q=0$ (when the number of the electrons in the flake is equal to the number of carbon atoms, what corresponds to the state of half-filling of the energy levels). In such a situation the zero-energy states lie at the Fermi level. Let us mention that in presence of the coulombic interaction between the electrons, it is energetically favourable to fill these states according to the Hund rule, so that at half-filling all the electrons occupying the zero-energy states have parallel spins. Therefore, the state with $\Delta q=0$ leads to a maximum spin of the nanoflake, equal to $(M-1)/2$ \cite{Wang2008,Ezawa2008b,Yazyev2010,Potasz2012}. Doping with either electrons (i.e. $\Delta q>0$) or holes ($\Delta q<0$) leads to decrease of the net magnetic moment, which vanishes for completely empty or completely filled shell of zero-energy states. Once the shell is either empty or completely filled, further doping may cause the magnetic moment to increase again. The dependence of the total spin of the nanoflake on the charge doping resulting from our Hamiltonian is illustrated for three nanoflakes, $M=2,3,4$, in Fig.~\ref{fig:fig1b}(a). Let us note that the doped triangular flakes were subject to extensive studies in \cite{Potasz2009,Potasz2012,Sheng2012}. 

The existence of the zero-energy states and the form of their wavefunctions can be expected to have a crucial impact on the indirect coupling between the magnetic moments in the nanoflakes. In particular, the nonzero spin distribution arising due to the presence of zero-energy states influences the mentioned properties. 

Let us illustrate briefly the spin distribution associated with the zero-energy states of zigzag triangular nanoflakes (for an example of the individual states in a nanoflake with $M=4$ see Fig.~5 in  \cite{EzawaBook}). In Fig.~\ref{fig:fig1b}(b), the total spin density (which corresponds to the difference between the charge density of spin up and spin down electrons) for the nanoflake with $M=3$ is shown for $\Delta q=0$ and in the limit of $U/t\to 0$. The radii of the circles correspond proportionally to the spin magnitude at the given lattice site while different colours depict the opposite signs of spin density. It is visible that the nonzero spin density is present only at the majority sublattice sites. Moreover, the wavefunctions of the zero-energy states are strongly localized at the zigzag edges of the nanoflake. The situation is changed when finite $U/t$ is introduced. For $U/t=1$ [Fig~\ref{fig:fig1b}(b)], the spin density takes the opposite signs at both sublattices, however, it is still dominant at the minority sublattice sites. The spin orientations at both sublattices are opposite. The dependence of the spin density at four representative lattice sites at the zigzag edge is plotted in Fig.~\ref{fig:fig1b}(c) for a wide range of the Hubbard exchange energies $U/t$. It is visible that the presence of coulombic interactions strongly promotes the spin polarization of both sublattices (of course conserving the total nanoflake spin). The increase rate of the polarization tends to drop for large $U/t$, reaching the value of a few, when the spin associated with each site reaches noticeably high values. Let us mention that a study of the magnetization of the finite graphene nanoflake as a function of the $U/t$ has been performed in \cite{Feldner2010} and a similar increase has been found within mean-field results. Comparison with exact diagonalization-based calculations shows that the rise of magnetization with Hubbard exchange energy is rather more linear-like. However, the results for $U/t<2$ obtained within mean-field approach are quite accurate (this range located the system below a critical value of $U/t$ for transition to Mott-Hubbard insulator). Let us explain also that there is no spontaneous spin polarization of the nanoflake strictly for $U/t=0$, unlike it happens in the limit of $U/t \to 0$, when the vanishing coulombic interactions still break the symmetry [Fig.~\ref{fig:fig1b}(b)]. However, when the symmetry is broken for example by the presence of a pair of magnetic impurities, the results for weak $U/t$ are convergent to the system behaviour for $U/t=0$. Therefore, in further presentation of the indirect coupling calculations distinguishing between the cases of $U/t=0$ and $U/t\to 0$ is not critical.

Then let us consider a nanoflake with a pair of localized magnetic impurity moments. In presence of these moments, the total Hamiltonian $\mathcal{H}$ of the system takes the form of $\mathcal{H}_{0}+\mathcal{H}_{imp}$. Here, $\mathcal{H}_{imp}$ is the Anderson-Kondo term for the interaction between the magnetic impurity spins and spins of the charge carriers, with exchange energy equal to $J$. For on-site impurities located at the sites $a,b$, the term is of the form $\mathcal{H}_{imp}=\frac{J}{2}S^{z}_{a}\left(n_{a,\uparrow}-n_{a,\downarrow}\right)+\frac{J}{2}S^{z}_{b}\left(n_{b,\uparrow}-n_{b,\downarrow}\right)$, while for plaquette impurities, positioned in the centers of the hexagons, it yields $\mathcal{H}_{imp}=\frac{J}{2}S^{z}_{a}\sum_{i\in NN }^{}{\left[\left(n_{i,\uparrow}-n_{i,\downarrow}\right)\right]}+\frac{J}{2}S^{z}_{b}\sum_{i\in NN}^{}{\left[\left(n_{i,\uparrow}-n_{i,\downarrow}\right)\right]}$ (where the summation is over all the sites belonging to the hexagon housing the impurity).

The total electronic energy $E_{tot}$ is then obtained for a nanoflake with $N+\Delta q$ electrons, separately for ferromagnetic (F) and antiferromagnetic (AF) orientation of the magnetic impurity spins. This allows for determination of an indirect exchange coupling RKKY-like integral from $2S^2J^{RKKY}=E_{tot}^{AF}-E_{tot}^{F}$ \cite{Annika2010a,Szalowski2011}. Therefore, the interaction between the impurity spins is projected onto an effective Hamiltonian of the form $H_{ab}=-J^{RKKY}\,S^{z}_{a}S^{z}_{b}$. The Ising form of Anderson-Kondo term leads to Ising-like interaction between impurity spins, however, for isotropic Heisenberg exchange the Hamiltonian $H_{ab}$ would be of the Heisenberg form with coupling energy $J^{RKKY}$ unchanged. Therefore the choice of the Ising form is only for the sake of facilitating the numerical calculations. Let us note that, due to preserved electron-hole symmetry of the total Hamiltonian, all the results are expected to be independent on the sign of $\Delta q$, i.e. identical for electron- and hole-doping. We also must emphasize that the present method yields the total energy of indirect coupling, which may involve mechanisms which differ from the usual RKKY interaction mechanism operative via electronic states occupied by pairs of opposite spin electrons. However, let us call the resulting coupling $J^{RKKY}$ for brevity.

\section{Numerical results}
\label{results}

\begin{figure}
  \includegraphics[scale=0.68]{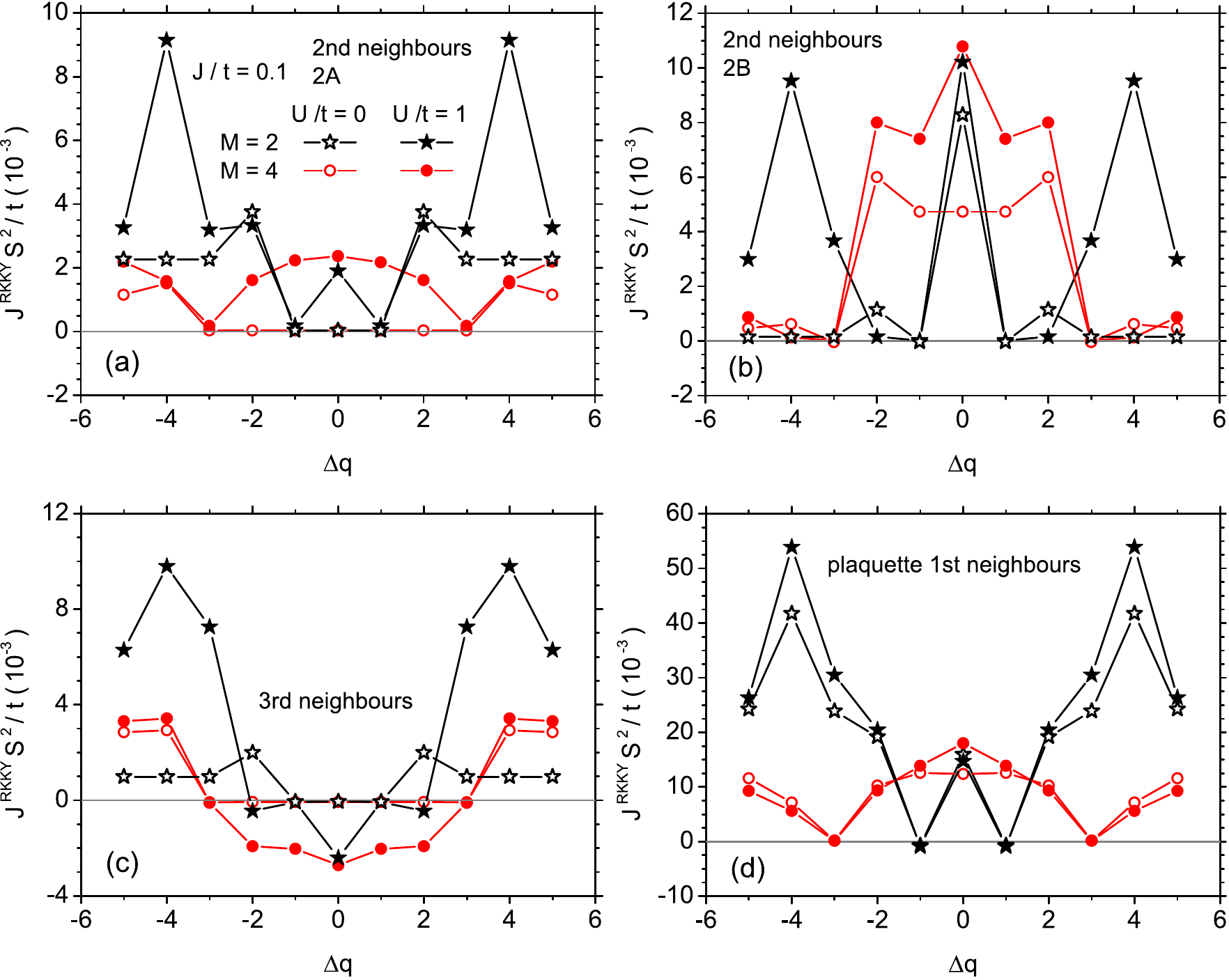}
\caption{Indirect coupling energy values as a function of charge-carrier doping for two selected nanoflakes, $M=2,4$, in absence ($U/t=0$) and in presence ($U/t=1$) of the Hubbard term. (a) 2A 2nd neighbours pair; (b) 2B 2nd neighbours pair; (c) 3rd neighbours pair; (d) 1st neighbours plaquette pair. For explanation see Fig.~\ref{fig:fig1}.}
\label{fig:fig4}
\end{figure}

The numerical results, presented in the subsequent figures were obtained by means of the total energy calculation, on the basis of exact diagonalization of the Hamiltonian (\ref{eq:eq1}) in the single particle approximation. For numerical calculations, the LAPACK package was used \cite{Lapack}. Let us note that these results are non-perturbational with respect to the $J$ parameter, describing the interaction between impurity spins and spins of the charge carriers, like in \cite{Szalowski2011,Annika2010a,Annika2010b}. This allows to capture the effects beyond the second-order perturbational calculus, the presence of which can be expected in such molecular-like systems; on the other hand, only the numerical results can be obtained. For the further studies, the value of $J/t=0.1$ was accepted (unless the $J$-dependence of coupling energy is studied).   

The triangular GNFs considered in the present study were characterized by the values of $M$ ranging between $M=2$ and 7. Several representative placements of the magnetic impurities were taken into consideration, strongly focusing on the position close to the zig-zag edge. For on-site impurities, two inequivalent second-neighbour pairs, 2A and 2B (see Fig.~\ref{fig:fig1}) were discussed (i.e. impurities in minority or in majority sublattice) in as well as the third-neighbour impurities placed in different sublattices. In addition, plaquette impurities (placed in the centres of the hexagons) were also considered in nearest-neighbour position. For studies of the distance dependence of the indirect coupling, also the other impurity positions were taken into account [as illustrated in Fig.~\ref{fig:fig6}(a)].

At the beginning, the importance of the Hubbard term for the indirect exchange integrals is verified. Fig.~\ref{fig:fig4} presents a comparison of the values of coupling energies obtained for $U/t=0$ and for $U/t=1$, the latter being the value accepted for the Hubbard term for further calculations. The dependence of the indirect coupling on the charge doping $\Delta q$ is demonstrated for two flakes characterized by $M=2$ and $M=4$. The results are plotted for 2A and 2B second-neighbour on-site impurities, for third-neighbour on-site impurities and for nearest-neighbour plaquette impurities, all located at the edge of the nanoflake. In general, it can be observed that for vanishing coulombic interactions, the indirect coupling is very weak for the cases of 2nd neighbours in minority sublattice or for 3rd neighbours, close to the charge neutrality. It remains negligible as long as the zero-energy states shell is partially filled (i.e., for $|\Delta q|<M-1$). On the contrary, for plaquette impurities and for 2nd neighbours in majority sublattice, the interaction has considerable magnitude in the vicinity of the charge neutrality point, for $|\Delta q|<M-1$. Including the Hubbard term with $U/t=1$ crucially influences the behaviour of an indirect coupling for 2A and 3rd neighbours position. Namely, the interaction energy is increased (toward ferromagnetic values for 2nd neighbours and antiferromagnetic values for 3rd neighbours). This effect remains robust against charge doping until $|\Delta q|<M-1$. For 2B and plaquette impurities, the effect is similar in its direction, however, much less pronounced. In all cases, the peak coupling energies correspond to $\Delta q=0$. Let us observe another peaks for indirect coupling present at $|\Delta q|=4$ for $M=2$, corresponding to another maximum nanoflake spin configuration [see Fig.~\ref{fig:fig1b}(a)].

\begin{figure}
\includegraphics[scale=0.68]{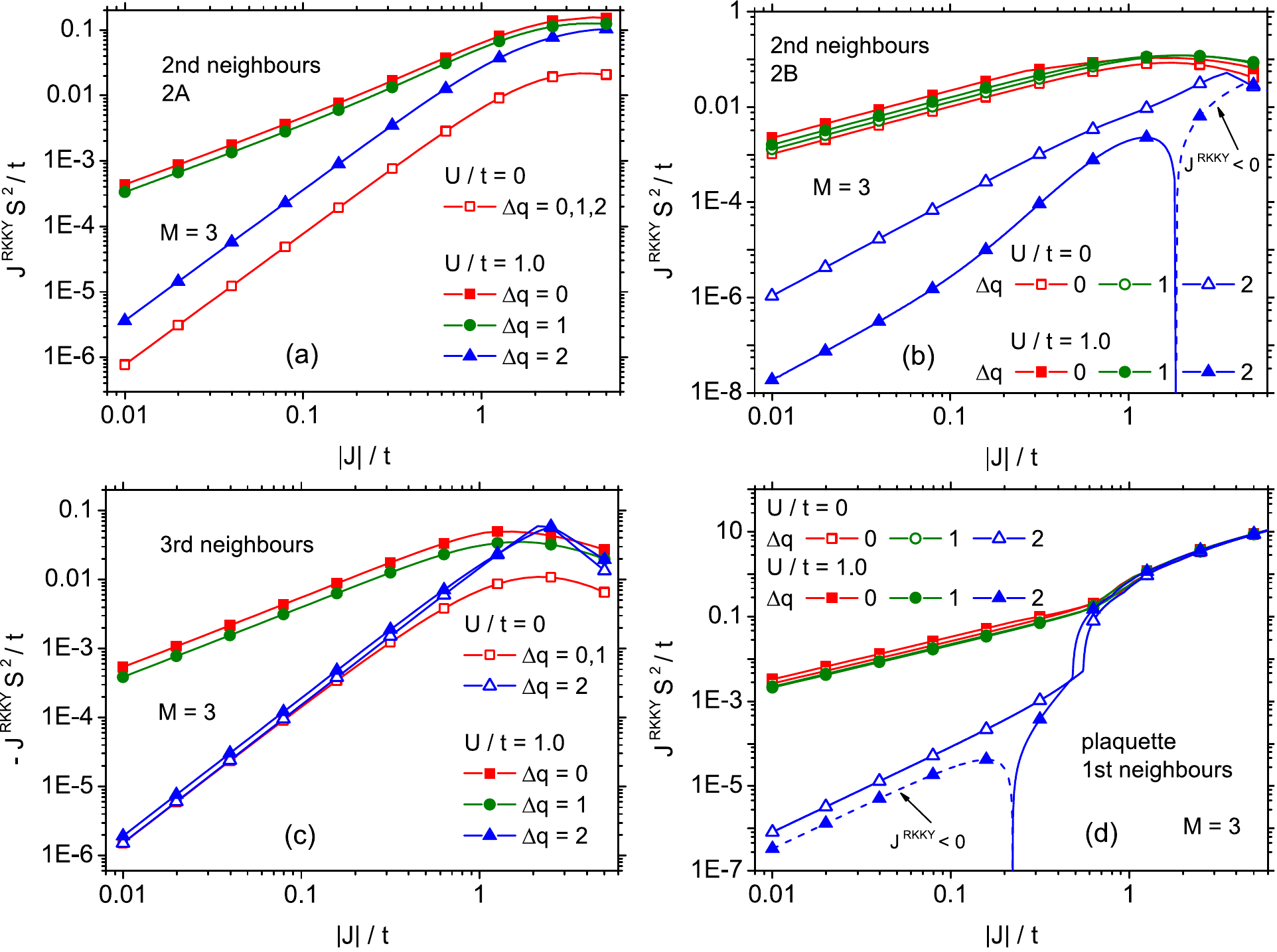}
\caption{Indirect coupling energy values as a function of the localized spin-electron spin exchange energy $J$, for charge neutrality $\Delta q=0$ and for two values of doping $|\Delta q|=1,2$, for a nanoflake with $M=3$. (a) 2A 2nd neighbours pair; (b) 2B 2nd neighbours pair; (c) 3rd neighbours pair; (d) 1st neighbours plaquette pair. For explanation see Fig.~\ref{fig:fig1}.}
\label{fig:fig2}
\end{figure}

In order to explain this behaviour, let us focus on the dependence of the indirect coupling energies on the exchange parameter $J$, describing the interaction between localized spin and electron spin. Such results are illustrated in Fig.~\ref{fig:fig2}, for $M=3$ and four impurity locations at the nanoflake edge, in doubly logarithmic scale. Our calculations rely on non-perturbative procedure of determining the total energy, thus are able to identify the contributions to an indirect coupling described by various orders of perturbational calculus. For $U/t=0$ and for 2A and 3rd neighbours, the coupling varies quadratically with $J$ and is weak in magnitude. This suggests the origin from the second-order perturbational process, which is the typical origin of indirect RKKY coupling. In the mentioned cases, the interaction is also weakly dependent on the charge doping. If $U/t=1$ is taken into consideration, the situation is changed. Then, the indirect coupling becomes a linear function of $|J|$ and the magnitude is greatly pronounced, if $|\Delta q|<2$, while its behaviour resembles the case of $U/t=0$ for $|\Delta q|=2$. Such a behaviour indicates, that a leading contribution originates form the first-order perturbational process, involving the zero-energy states, the shell of which is partially filled for $|\Delta q|<2$.  On the other hand, for 2B neighbours and plaquette impurities, both for $U/t=0$ and 1 the linear dependence of indirect coupling on $|J|$ is observed for $|\Delta q|<2$, and again, for empty shell of the zero-energy states the usual quadratic behaviour is restored. As it can be deduced form the Fig.~\ref{fig:fig1b}(b) and (c), nonzero spin density coming from a half-filled shell of the zero-energy states is present only at the majority sublattice for $U/t\to 0$, and switching on finite $U/t$ causes the redistribution of the spin density to both sublattices. Therefore, for $U/t\to 0$, the first-order perturbational contribution is possible only for the impurities in the majority sublattice (or for plaquette impurities, each of which interacts with equal number of electrons residing at both sublattices). Increase of $U/t$ enables the same mechanism for impurity pairs in minority sublattice (e.g. 2A) or for impurities in both sublattices (e.g. 3rd neighbours). 

\begin{figure}
\includegraphics[scale=0.68]{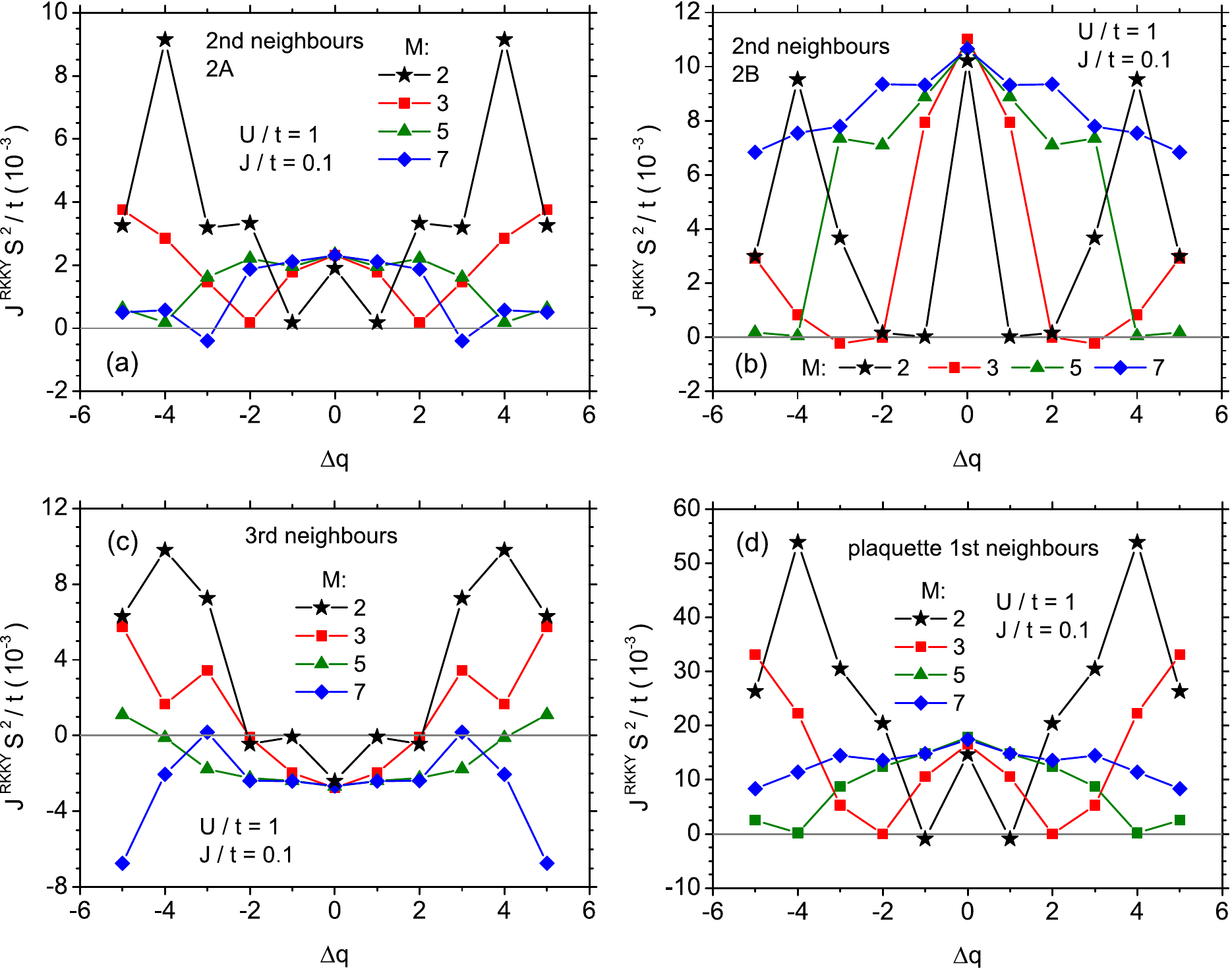}
\caption{Indirect coupling energy values as a function of charge-carrier doping for increasing size of the nanoflake $M=2,3,5,7$, in presence of the Hubbard term ($U/t=1$). (a) 2A 2nd neighbours pair; (b) 2B 2nd neighbours pair; (c) 3rd neighbours pair; (d) 1st neighbours plaquette pair. For explanation see Fig.~\ref{fig:fig1}.}
\label{fig:fig3}
\end{figure}

Let us consider the first-order perturbational mechanism quantitatively \cite{Szalowski2011}. The first-order correction to energy of all the states due to presence of impurity spin $S_{a}$ in site $a$ is $\Delta E=-|J|S^{z}_{a}s^{z}_{a}$ and is proportional to the total spin in this site $s^{z}_{a}=\left(n_{a,\uparrow}-n_{a,\downarrow}\right)/2$. Therefore, if the spin-degenerate states are filled with pairs of opposite-spin electrons and no net spin polarization is present, the first order correction vanishes. Let us assume that two impurity spins are present at the sites $a$ and $b$ (assuming $|s^{z}_{a}|\geq |s^{z}_{b}|$). If the impurity spins are polarized ferromagnetically, the total energy change is $\Delta E^{F}_{ab}=-|J|Ss^{z}_{a}-\eta|J|Ss^{z}_{b}$, where $\eta=s^{z}_{a} s^{z}_{b}/\left|s^{z}_{a} s^{z}_{b}\right|$ is the sign of $s^{z}_{a} s^{z}_{b}$. The orientation of the spins $s^{z}_{a}$ and $s^{z}_{b}$ depends on the sign of $J$ such that it maximizes the absolute value of the correction (which itself is negative), of course conserving the relative orientation of $s^{z}_{a}$ and $s^{z}_{b}$. On the other hand, for antiferromagnetic orientation of impurity spins, it yields $\Delta E^{AF}_{ab}=-|J|Ss^{z}_{a}+\eta |J|Ss^{z}_{b}$. The indirect coupling energy due to this mechanism is then $2S^2 J^{RKKY}= \Delta E^{AF}_{ab}-\Delta E^{F}_{ab}=2\eta S|J|s^{z}_{b}$. Let us note that owing to linear dependence on $|J|$, it dominates over the usual $J^2$-proportional mechanism. 

The idea of the first-order perturbational mechanism can be expressed briefly by the statement that the energetically favourable mutual orientation of impurity spins follows the mutual orientation of the electron spin at sites $a$ and $b$. Therefore, it is ferromagnetic for impurities in the same sublattice and antiferromagnetic in different sublattices [see Fig.~\ref{fig:fig1b}(b)]. 

In Fig.~\ref{fig:fig2} it is visible, that for strong exchange coupling between impurity spins and electron spins $J/t$, taking the value of a few, the coupling energy tends to saturate, and then even decrease; also the change of sign is possible in some cases, as for 2B 2nd neighbours and plaquette impurities, in presence of Hubbard term and for empty zero-energy shell. 

The evolution of an indirect coupling with the size of triangular nanoflake can be analysed in Fig.~\ref{fig:fig3}, where the dependence of coupling energy on charge doping is presented for $M=2,\dots,7$. The results take into account the value of $U/t=1$. In all the presented cases, it is visible that the coupling depends in principle rather weakly on charge doping as long as $|\Delta q|<M-1$, i.e. as long as the shell of zero-energy states is neither completely empty nor completely filled. The coupling energy in general gradually decreases with doping within mentioned range. The situation is slightly different only for $M=7$, where for 2A and 3rd neighbours, the coupling diminishes at $|\Delta q|=3$. It can be concluded that a quite strong indirect coupling, robust against charge doping, can be achieved at the zigzag edge of the triangular nanoflake.

\begin{figure}
  \includegraphics[scale=0.68]{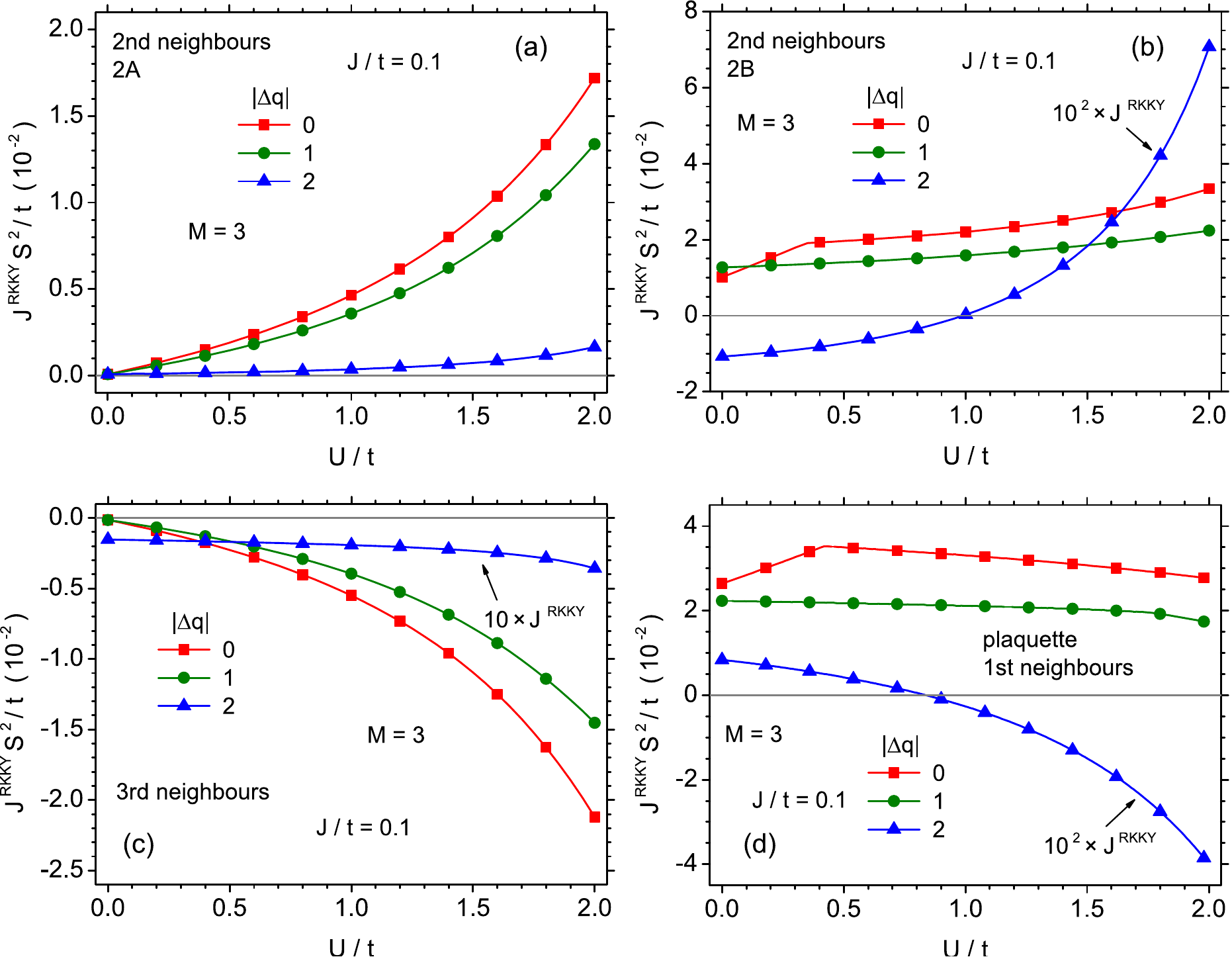}
\caption{Indirect coupling energy values as a function of the Hubbard on-site exchange energy $U/t$, for charge neutrality $\Delta q=0$ and for two values of doping $|\Delta q|=1,2$, for a nanoflake with $M=3$. (a) 2A 2nd neighbours pair; (b) 2B 2nd neighbours pair; (c) 3rd neighbours pair; (d) 1st neighbours plaquette pair. For explanation see Fig.~\ref{fig:fig1}.}
\label{fig:fig5}
\end{figure}

\begin{figure}
  \includegraphics[scale=0.90]{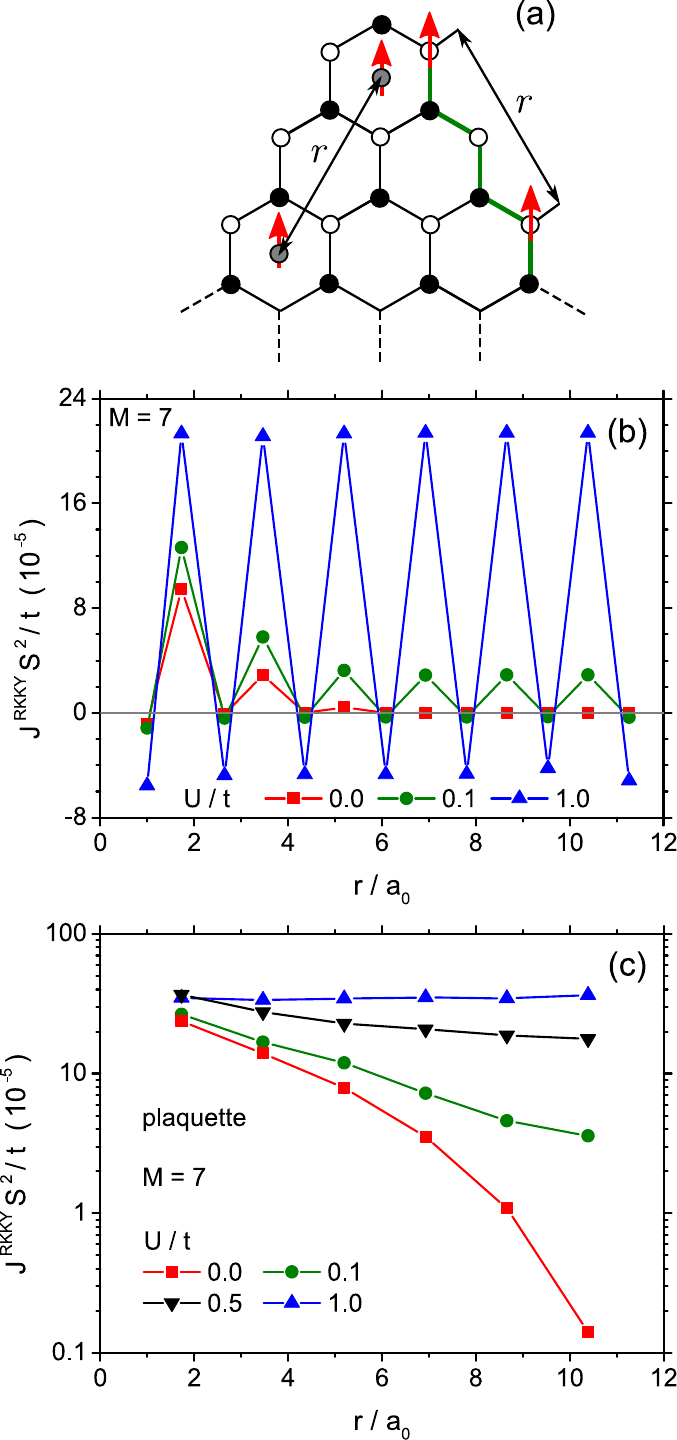}
\caption{(a) distance between pairs of on-site impurities and plaquette impurities along the zigzag edge; (b) Indirect coupling energy values as a function of the distance between the on-site impurities for a nanoflake with $M=7$ for various strengths of Hubbard on-site exchange energy $U/t$; (c) Indirect coupling energy values as a function of the distance between the plaquette impurities for a nanoflake with $M=7$ for various strengths of Hubbard on-site exchange energy $U/t$.}
\label{fig:fig6}
\end{figure}

The effect of the Hubbard exchange energy parameter $U$ on the indirect coupling energy can be followed in the Fig.~\ref{fig:fig5}, where the cases of two 2nd neighbour impurities as well as 3rd neighbours and plaquette impurities are studied. The state of the charge neutrality and moderate dopings ($|\Delta q|=1,2$) are considered. For 2nd neighbours 2A [Fig.~\ref{fig:fig5}(a)] and for 3rd neighbours [Fig.~\ref{fig:fig5}(c)], the coupling energy increases noticeably with the increase of $U$, strengthening the interaction over orders of magnitude. This is due to the mentioned fact, that for $U/t= 0$, there is no first-order contribution to indirect coupling for these impurity pairs (as the spin density vanishes at the minority sublattice under such circumstances). As Hubbard exchange is switched on, the opposite-sign spin density develops at both sublattices [see Fig.~\ref{fig:fig1b}(b) and (c)] and first-order contribution (of ferro- or antiferromagnetic kind) appears and increases in strength. When a zero-energy states shell is empty ($|\Delta q|=2$), the changes in indirect coupling energy are much less pronounced. The situation is somehow different for 2B 2nd neighbours and plaquette impurities [Fig.~\ref{fig:fig5}(b) and (d), respectively], where first-order contribution to indirect coupling exists already at $U/t= 0$, so that the increase in Hubbard exchange energy only moderately influences the interaction between impurities for $|\Delta q|<2$. For empty zero-energy states shell, the indirect coupling tends to change sign when $U/t$ increases, which can be attributed to charge density redistribution. This happens also under the influence of increasing $J/t$ [see Fig.~\ref{fig:fig2}(b)]. Note that the coupling magnitude in these cases is greatly reduced in comparison with 2A 2nd neighbours or 3rd neighbours. 

It should be emphasized that for plaquette impurities in undoped graphene monolayer, their indirect interactions are expected to be antiferromagnetic \cite{Annika2010a}. However, at the edge of the studied nanoflakes, the coupling has ferromagnetic sign for half-filling conditions and for low charge dopings, as shown in Fig.~\ref{fig:fig5}(d). This is mainly due to the influence of the first-order coupling mechanism. The spin densities originating from zero-energy states are opposite in sign, but of unequal absolute values for both sublattices [see Fig.~\ref{fig:fig1b}(b)], so that a nonzero net spin polarization is present at a given hexagon, being especially pronounced for locations close to the edge. Moreover, the orientation of net spin for each hexagon is the same. Therefore, a ferromagnetic coupling is established between plaquette impurities under such conditions. This constitutes an important finite-size effect in the indirect coupling in the studied structures.

Finally, let us study the distance dependence of an indirect coupling along the zigzag edge of an undoped nanoflake. First we focus on the on-site magnetic impurities and for the purpose of this analysis, one of the spins is located in majority sublattice site closest to the edge, while the other takes various positions along the zigzag edge, starting from nearest neighbours [as illustrated in Fig.~\ref{fig:fig6}(a)]. The calculations are performed for a nanoflake with $M=7$. The results are depicted as a function of distance in Fig.~\ref{fig:fig6}(b) for three values of $U/t$: 0, 0.1 and 1. It is visible that without coulombic interactions included, the coupling tends to vanish considerably fast (it can be verified that the decay is approximately exponential in character or even faster). Its sign oscillates according to the rule that coupling is ferromagnetic for on-site impurities in the same sublattice while it is antiferromagnetic for different sublattices for half-filling conditions. Inclusion of the Hubbard term significantly changes the distance behaviour of an indirect coupling. For $U/t=0.1$, the coupling magnitudes are enhanced and the distance dependence of coupling becomes rather weak after some initial fall. For $U/t=1$, the coupling magnitude becomes almost distance-independent (regarded separately for the same and different sublattices) and the sign oscillates between the two cases. This reflects the dominance of the discussed coupling mechanism based on zero-energy states. For significant $U/t$ value, the coupling energies are weakly dependent on the distance owing to almost uniform distribution of spin density values along the edge for both sublattices [see Fig.~\ref{fig:fig1b}(b)].

Secondly, we study the distance dependence for the case of plaquette impurities located along the edge [see Fig.~\ref{fig:fig6}(a)]. The results are presented in Fig.~\ref{fig:fig6}(c) in logarithmic scale. The coupling is always ferromagnetic along the edge and in absence of Hubbard term it decays exponentially with the distance for small distances, while the decay becomes even faster for larger separation of impurities. Switching on the coulombic interactions enhances the coupling energy and flattens the distance dependence. Like for on-site impurities, for $U/t=1$, the coupling loses its distance dependence along the zigzag edge. Once more, let us mention that in this case the coupling sign is ferromagnetic, while the opposite sign is predicted for infinite graphene monolayer.

Let us note that an influence on the coulombic interactions taken into account by means of a Hubbard term has been studied for monolayer graphene and zigzag nanoribbons by Black-Schaffer \cite{Annika2010b}. In particular, a distance-independent coupling at zigzag edge of a narrow zigazg nanoribbon (which indicated spontaneous magnetic polarization at the edge) has been found in accordance with the present findings. Moreover, also the mechanism of a first-order contribution to the coupling, proportional to the electron spin at one of the impurity sites has been identified in such case.

\section{Concluding remarks}
\label{conclusion}

In the paper, the indirect coupling between localized on-site and plaquette magnetic moments on the triangular graphene nanoflakes with zig-zag edge was studied. Particular attention was paid to the role of a shell of zero-energy states present in these nanostructures. In that context, the influence of charge coupling has been studied and the importance of Hubbard on-site exchange term was emphasized. In the presence of edge-localized zero-energy states, the placement of the impurities at the nanoflake edge yields particularly interesting behaviour of an indirect coupling. Let us note that the edge-localized states for example in pits in graphene layer have been observed by means of STM/STS methods already in \cite{Klusek2001,Klusek2005}. Due to the edge-localized form of the wavefunction of the zero-energy states and nonzero spin density at the ground state for charge neutrality, a first-order perturbational mechanism becomes operative for many impurity positions along the edge. This kind of an indirect coupling, similar to a double exchange, dominates strongly over the usual RKKY second-order perturbative mechanism in numerous cases. Depending on the on-site impurity location, the enhanced indirect coupling can be either ferromagnetic (for impurities in the same sublattice) or antiferromagnetic (impurities in various sublattices). If the exchange energy between localized on-site spins and electron spins is not too large, we find no deviation from this sign rule related to RKKY coupling in graphene for on-site impurities. The situation is different for plaquette impurities. While in undoped graphene monolayer the RKKY coupling between such impurities is antiferromagnetic \cite{Annika2010a}, our results show that it becomes ferromagnetic for impurities located at the edge of a triangular nanoflake. This sign change is attributed mainly to the first-order coupling mechanism.  

The partially filled shell of zero-energy states guarantees the coupling which is quite robust against charge doping in larger nanoflakes; on the other hand, in the smallest structures it can be switched form large value to a small one by doping with a single charge carrier only.  

The distance dependence of the coupling along the edge is also heavily influenced by the presence of the coulombic on-site interactions. Namely, the fast decay of the coupling for $U/t=0$ is replaced with almost flat distance dependence (distinguishing only between the sublattices) for $U/t=1$. 

The obtained results for triangular nanoflakes demonstrate that the indirect coupling between impurity spins in such system shows a nontrivial dependence on the charge doping and impurity position. This properties originate mainly from a subtle interplay between first- and second-order perturbative contributions. Such a rich behaviour may motivate further studies of routes to control the coupling. Moreover, it might also stimulate further works on spin readout in such molecular magnet-like systems (see e.g. \cite{Candini2011}).

\section{Acknowledgments}
This work has been supported by Polish Ministry of Science and Higher Education on a special purpose grant to fund the research and development activities and tasks associated with them, serving the development of young scientists and doctoral students.

The computational support on HUGO cluster at Department of Theoretical Physics and Astrophysics, P. J. \v{S}af\'{a}rik University in Ko\v{s}ice is gratefully acknowledged.

The author is deeply grateful to T. Balcerzak for critical reading of the manuscript.



\bibliographystyle{elsarticle-num}







\end{document}